\documentclass{jfm}
\usepackage{color}
\usepackage{amstext}
\usepackage{amssymb}
\usepackage{graphicx}
\usepackage{amsmath}

\newcommand\curl{\mathrm{curl} \,}

\newcommand\const{\mathrm{const}}
\newcommand\Div{\mathrm{div}\,}

\newcommand\vF{\boldsymbol{F}}

\newcommand\vf{\boldsymbol{f}}

\newcommand\vu{\boldsymbol{u}}

\newcommand\vx{\boldsymbol{x}}

\newcommand\vomega{\boldsymbol{\omega}}

\newcommand\valpha{\boldsymbol{\alpha}}
\newcommand\vbeta{\boldsymbol{\beta}}

\begin{document}

{\title[Fluid Flows driven by Oscillating Body Force] {Fluid Flows driven by Oscillating Body Force}}

\author[V. Vladimirov and N. Peake]
{V.\ns A.\ns V\ls l\ls a\ls d\ls i\ls m\ls i\ls r\ls o\ls v\ns and\ns N.\ns  P\ls e\ls a\ls k\ls e}

\affiliation{Sultan Qaboos University, University of York,  and University of Cambridge}


\setcounter{page}{1}\maketitle \thispagestyle{empty}

\begin{abstract}

In this note we consider general formulation of Euler's equations for an inviscid incompressible homogeneous fluid with an oscillating body force. Our aim is to derive the averaged equations for these flows with the help of two-timing method. Our main result is the general and simple form of the equation describing the averaged flows, which are derived without making any additional assumptions.
The presented results can have many interesting applications.

\end{abstract}

\section{Introduction \label{sect01}}

Oscillating flows play a central role in fluid dynamics, their key appearances in various applications in medicine, biophysics, geophysics,
engineering, astrophysics, acoustics are well known. In this note we consider inviscid oscillating incompressible flows driven by an oscillating in time body force (in particular, it can be a rotating body force). For doing that we use the two-timing method. Our main result is the deriving of a general and simple form of the averaged Euler's equations by the two-timing method.
The used two-timing technique is taken in the same form as in \cite{Vladimirov0, Yudovich, Vladimirov1, Vladimirov, VladimirovL, VladProc}.

There are many attempts to describe the fluid flows with very complex boundary conditions by the replacing of moving and deforming boundaries by various body forces. These research directions are motivated by important applications in such areas as turbo-machinery, biological and medical fluid dynamics.
The most popular approaches here are the penalization method and the immersed boundary methods, see
\cite{Peskin1, Peskin2, Schnider, Mittal, Peake}. However it is surprising that until now the averaged equations of flows cased by oscillating body forces has escaped any attention of researchers.

\section{Two-timing problem and used notations}

We study the dynamics of a homogeneous inviscid incompressible fluid with velocity field $\vu^*$
and vorticity $\vomega^*\equiv\nabla^*\times\vu^*$ (asterisks mark dimensional variables). The governing
equations in cartesian coordinates $\vx^*=(x_1^*,x_2^*,x_3^*)$ and time ${t}^*$ are
\begin{eqnarray}\label{exact-1}
&&{\partial\vu^*}/{\partial {t}^*}+(\vu^*\cdot\nabla^*)\vu^*=-\nabla^* p^*+\vf^*(\vx^*,\tau),\quad \nabla^*\cdot\vu^*=0
\end{eqnarray}
where $\nabla^*=(\partial/\partial x_1^*, \partial/\partial x_2^*,\partial/\partial x_3^*)$  and $\vf^*(\vx^*,\tau)$ is a given external body force being a periodic function of variable $\tau=\sigma^*t^*$, where $\sigma^*$ is a given frequency. For brevity we include the constant density into $p^*$ and $\vf^*$. Also for simplicity we accept that the fluid fills all three dimensional space.

We accept that the considered class of oscillating
flows possesses the characteristic scales of velocity $U$, length $L$, and frequency $\sigma^*$
\begin{eqnarray}
&& U,\quad L,\quad  \sigma^*;\quad T\equiv L/U
\label{scales-list}
\end{eqnarray}
where $T$ is a dependent time-scale. The dimensionless (not asteriated) variables and frequency are
\begin{eqnarray}
&& \vx\equiv\vx^*/L,\quad t\equiv t^*/T,
\quad{\vu}\equiv{\vu}^*/U,\quad\sigma\equiv\sigma^*T\gg 1
\label{scales}
\end{eqnarray}
where $1/\sigma$ is the small parameter of our asymptotic theory.
In the dimensionless variables \eqref{exact-1} takes place
\begin{eqnarray}\label{exact-1a}
&&{\partial\vu}/{\partial {t}}+(\vu\cdot\nabla)\vu=-\nabla p + \vf(\vx,\tau)
\end{eqnarray}
Here and below we do not write the condition $\Div\vu=0$, but always keep it in mind.
We consider the solutions of \eqref{exact-1a} in the two-timing form with two time-variables
\begin{eqnarray}
&&\tau\equiv\sigma {t},\, s\equiv t\label{exact-2}
\end{eqnarray}
where $s$ and $\tau$ are two \emph{mutually dependent} time-variables (we call $s$ \emph{slow time} and $\tau$
\emph{fast time}).
Then the use of the chain rule brings \eqref{exact-1a} to the form
\begin{eqnarray}
&&\sigma{\vu}_\tau+ \vu_s + (\vu\cdot\nabla)\vu=-\nabla p + \vf(\vx,\tau),
\quad \varepsilon\equiv 1/\sigma\to 0\label{exact-2a}
\end{eqnarray}
where the subscripts $\tau$ and $s$ stand for the partial derivatives.

The key suggestion of the two-timing method is
\begin{eqnarray}
&&\emph{$\tau$ and $s$ are considered as mutually independent variables}\label{key}
\end{eqnarray}
As a result, we convert  \eqref{exact-2a} from a PDE with independent variables $t$ and $\vx$ into a PDE with the extended number of independent variables $\tau, \, s$ and $\vx$. Then the solutions of \eqref{exact-1a} must have a functional form:
\begin{eqnarray}
&& \vu={\vu}(\vx, s, \tau)\label{exact-3}
\end{eqnarray}
It should be emphasized, that without \eqref{key} a functional form of solutions can be different from \eqref{exact-3}; indeed the presence of the dimensionless scaling parameter $\varepsilon$ allows one to build an infinite number of different time-scales, not just $\tau$ and $s$. In this paper we accept \eqref{key} and analyse the related averaged equations and solutions in the functional form \eqref{exact-3}.

To make further analytic progress, we introduce few convenient notations and agreements.
Here and below we assume that \emph{any dimensionless function} $g(\vx,s,\tau)$ has the following properties:

(i)  $g\sim {O}(1)$ and  all its required for consideration $\vx$-, $s$-, and $\tau$-derivatives are also ${O}(1)$;

(ii) $g$  is $2\pi$-periodic in $\tau$, \emph{i.e.}\ $g(\vx, s, \tau)=g(\vx,s,\tau+2\pi)$ (about this technical simplification see the Discussion section);

(iii) $f$ has an average given by
\begin{equation}\label{aver}
\overline{g}\equiv \langle {g}\,\rangle \equiv \frac{1}{2\pi}\int_{\tau_0}^{\tau_0+2\pi}
g(\vx, s, \tau)\, d \tau \qquad \forall\ \tau_0=\const;
\end{equation}

(iv) $g$ can be split into averaged and purely oscillating parts
\begin{equation}
g(\vx, s, \tau)=\overline{g}(\vx, s)+\widetilde{g}(\vx, s, \tau)\label{decompos}
\end{equation}
where  \emph{tilde-functions} (or  purely oscillating functions) are such that $\langle \widetilde g\, \rangle =0$ and the \emph{bar-functions} $\overline{g}(\vx,s)$ (or the averaged functions $\langle {g}\,\rangle=\overline{g}$) are $\tau$-independent;

(v)  we introduce a special notation $\widetilde{g}^{\tau}$ (with a superscript $\tau$) for  the
\emph{tilde-integration} of tilde-functions, such integration keeps the result in the tilde-class. For doing that we notice that the integral of a tilde-function
\begin{equation}\nonumber
G(\vx,s,\tau)\equiv\int_0^\tau \widetilde{g}(\vx,s,\tau')\, d \tau'
\label{ti-integr0}
\end{equation}
often does not belong to the tilde-class. In order to keep the result of integration in the tilde-class we should subtract the average
\begin{equation}
\widetilde{g}^\tau\equiv G-\overline{G}
\label{ti-integr0}
\end{equation}
The tilde-integration is
inverse to the $\tau$-differentiation $({{\widetilde{g}}}^{\tau})_{\tau}=({g}_{\tau})^{\tau}={\widetilde{g}}$; the
proof is omitted.

Here we emphasize that in all text below all large or small parameters are represented by various degrees of
$\sigma$ only; these parameters appear as explicit multipliers in all formulae containing tilde- and
bar-functions; while these functions are always of order one.

\section{Main equation and successive approximations \label{sect04}}

Let us  make some amplitude specification in \eqref{exact-2a}
\begin{eqnarray}
&&\sigma{\vu}_\tau+ \vu_s + (\vu\cdot\nabla)\vu=-\sigma\nabla p + \sigma\widetilde{\vf}(\vx,\tau)\label{exact-2b}
\end{eqnarray}
where we have chosen the magnitudes of $p$ and $\vf$ in terms of $\sigma$. Mathematically, this choice is required for  balancing the first term $\sigma{\vu}_\tau$ in the equation with  other terms of the same order. Physically, the choice of the force as $\sigma \vf$ is dictated by the fact that the flow is driven by this force, so, it has to be of the highest available order of magnitude.

We have also made a simplifying suggestion (just for this note): the given body force is chosen as being a purely oscillatory function with a zero mean $\vf=\widetilde{\vf}$ \eqref{decompos}. The explicit introduction of $\varepsilon\equiv1/\sigma$ converts \eqref{exact-2b} into
\begin{eqnarray}
&&{\vu}_\tau+ \varepsilon\vu_s + \varepsilon(\vu\cdot\nabla)\vu=-\nabla p + \widetilde{\vf}(\vx,\tau)\label{main-eq}
\end{eqnarray}
We are looking for the solutions
in the form of regular series
\begin{eqnarray}
&&(\vu,p)=\sum_{k=0}^\infty \varepsilon^k (\vu_k,p_k)\quad k=0,1,2,3\dots
\label{basic-4aa}
\end{eqnarray}
The substitution of
(\ref{basic-4aa}) into (\ref{main-eq}) produces the equations for successive approximations.

\emph{The equations of zero approximation} of (\ref{main-eq}) are
\begin{eqnarray}
&&{{\vu}}_{0\tau}=-\nabla p_0 + \widetilde{\vf}(\vx,\tau)\label{0-approx}\\
&&\Div\vu_0=0\nonumber
\end{eqnarray}
The bar-parts of these equations give us
\begin{eqnarray}
&&\overline{p}_0 =\overline{p}_0(s),\quad \Div\overline{\vu}_0=0\label{0-approx-bar}
\end{eqnarray}
while the tilde-parts lead to the conclusion
\begin{eqnarray}
&&{\widetilde{\vu}}_{0}=-\nabla \widetilde{p}_0^\tau + \widetilde{\vf}^\tau\label{0-approx-tilde}\\
&&\Div\widetilde{\vu}_0=0\nonumber
\end{eqnarray}
Taking the divergence of the first equation we obtain the Poisson equation for $\widetilde{p}_0^\tau$
\begin{eqnarray}
&&\Delta\widetilde{p}_0^\tau\equiv\nabla^2 \widetilde{p}_0^\tau = \Div\widetilde{\vf}^\tau\label{poisson}
\end{eqnarray}
which can be solved provided the boundary conditions at infinity are given. This solution can be symbolically written as:
\begin{eqnarray}
&& \widetilde{p}_0^\tau = \Delta^{-1}\left(\Div\widetilde{\vf}^\tau\right)\label{poissonA}
\end{eqnarray}
The boundary conditions at infinity can be chosen as $\widetilde{p}_0\to 0$ as $|\vx|\to\infty$. At the same time we can consider the class of functions $\widetilde{\vf}$ which are
rapidly decaying as $|\vx|\to \infty$ or  $\widetilde{\vf}\equiv 0$ outside a desired finite domain (say, outside the domain, modelling an oscillating heart or rotating turbine).
After solving \eqref{poissonA}, the first equation \eqref{0-approx-tilde} gives us the expression for ${\widetilde{\vu}}_{0}$.

\emph{The equations of the first approximation} of (\ref{main-eq}) are
\begin{eqnarray}
&&{\vu}_{1\tau}+ \vu_{0s} + (\vu_0\cdot\nabla)\vu_0=-\nabla p_1 \label{1-approx}\\
&&\Div{\vu}_1=0\nonumber
\end{eqnarray}
The bar-part of the first equation is
\begin{eqnarray}
&&\overline{\vu}_{0s} + (\overline{\vu}_0\cdot\nabla)\overline{\vu}_0+ \langle(\widetilde{\vu}_0\cdot\nabla)\widetilde{\vu}_0\rangle =-\nabla \overline{p}_1 \label{1-approx-bar}
\end{eqnarray}
where the already known function $\widetilde{\vu}_0$ is to be substituted from \eqref{0-approx-tilde}, \eqref{poissonA}. Using the identity
$$
({\vu}_0\cdot\nabla){\vu}_0=\vomega_0\times\vu_0+\nabla{\vu^2_0}/{2}
$$
we transform \eqref{1-approx-bar} into the final system of equations
\begin{eqnarray}
&&\overline{\vu}_{0s} + (\overline{\vu}_0\cdot\nabla)\overline{\vu}_0+ \langle(\curl\widetilde{\vf}^\tau)\times(\widetilde{\vf}^\tau-\nabla \widetilde{p}_0^\tau)\rangle =-\nabla {\overline{p}}_m \label{eqn-final}\\
&&\Div{\overline{\vu}}_0=0\nonumber
\end{eqnarray}
where $\overline{p}_m$ is a modified pressure. One can see that the resulting form of the averaged equations coincides with the standard Euler's equations containing an additional body force
\begin{eqnarray}\label{force}
\overline{\vF}=-\langle(\curl\widetilde{\vf}^\tau)\times(\widetilde{\vf}^\tau-\nabla \widetilde{p}_0^\tau)\rangle
\end{eqnarray}
where the oscillatory pressure represents the solution of Poisson's equation \eqref{poissonA}.
This formula shows that for $\overline{\vF}\neq 0$ we have to consider  body forces with $\curl\widetilde{\vf}\neq 0$.

\underline{\emph{Example 1:}} A simple example can be chosen as
\begin{eqnarray}\label{force2}
\widetilde{\vf}(\vx,s,\tau)= \overline{\valpha}(\vx)\sin\tau+ \overline{\vbeta}(\vx)\cos\tau
\end{eqnarray}
with new given functions $\overline{\valpha}(\vx)$ and ${\overline{\vbeta}}(\vx)$.
The averaged force $\overline{\vF}$ for this case can be calculated as
\begin{eqnarray}\label{force3}
\overline{\vF}=\frac{1}{2}\Big(\curl\overline{\valpha}\times(\overline{\valpha}-\nabla \overline{A})+\curl\overline{\vbeta}\times(\overline{\vbeta}-\nabla \overline{B})\Big)
\end{eqnarray}
where
$$
\overline{A}=\Delta^{-1}(\Div\overline{\valpha}), \quad \overline{B}=\Delta^{-1}(\Div\overline{\vbeta})
$$
\underline{\emph{Example 2:}} One can see that the force \eqref{force2} can be chosen as solenoidal with
$$
\Div{\widetilde{\vf}}=\Div\overline{\valpha}=\Div\overline{\vbeta}=\overline{A}=\overline{B}=0
$$
In this case $\widetilde{p}\equiv 0$ due to the uniqueness of solution to Laplace equation \eqref{poissonA}, which means $\widetilde{\vu}=\widetilde{\vf}_0^\tau$ and
\begin{eqnarray}\label{force4}
\overline{\vF}=\frac{1}{2}\Big((\overline{\valpha}\cdot\nabla)\overline{\valpha}+(\overline{\vbeta}\cdot\nabla)\overline{\vbeta}\Big)
\end{eqnarray}
where the gradient term is included to the modified pressure.
The expressions \eqref{force3} and \eqref{force4} can be further specified by the particular choice  of $\overline{\valpha}$ and $\overline{\vbeta}$.

\underline{\emph{Example 3:}} We can take $\overline{\vbeta}\equiv 0$, $\overline{\valpha}=(0, 0,\overline{\alpha}(r,z))$ in the cylindrical coordinates $(z, r,\phi)$, where $\overline{\alpha}(r,z)$ represents an arbitrary  smooth function. In this case \eqref{force4}  gives
\begin{eqnarray}\label{force5}
\overline{\vF}=\frac{1}{2}((\overline{\valpha}\cdot\nabla)\overline{\valpha})=(0, \overline{\alpha}^2/r,0)
\end{eqnarray}
and one can see that the oscillatory force with the only nonzero  angular component produces a radially directed averaged force.

\underline{\emph{Example 4:}} The force $\widetilde{\vf}$ \eqref{force2}  could have some relations to applications. Two terms are chosen in order to consider a  rotating force.
A simpler way of doing that is to consider three components of $\widetilde{\vf}$ in cylindrical $(z,r,\phi)$ or spherical $(r, \theta,\phi)$ coordinates, where all the components are functions of the form $\widetilde{f}(z,r, n\phi-\tau)$ or $\widetilde{f}(z,\theta, n\phi-\tau)$ with an integer $n$ and the azimuthal angle $\phi$.

This research is partially supported by
the grant IG/SCI/DOMS/16/13 from Sultan Qaboos University, Oman.

\section{Discussion}

1. A given force with non-zero average part can be routinely included into the above consideration. In this case one should take $\overline{\vf}+\sigma\widetilde{\vf}$ instead of $\sigma\widetilde{\vf}$ in \eqref{exact-2a} or an additional term $\varepsilon \overline{\vf}$ in the right hand side of \eqref{main-eq}. It will only case the appearance of an additional term in \eqref{force} as
\begin{eqnarray}\label{force1}
\overline{\vF}=
\overline{\vf}+\langle(\curl\widetilde{\vf}^\tau)\times(\widetilde{\vf}-\nabla \widetilde{p}_0)\rangle
\end{eqnarray}

2.  Viscosity can be included into consideration, however the form of resulting averaged equations will depend on the order of magnitude of Reynolds number in terms of $\varepsilon$. In particular, if $1/Re\sim\varepsilon$ then the standard `viscous' term can be just added to the equation \eqref{eqn-final}. 

3. The $\tau$-periodicity represents a functional restriction, which can be generalised as it is accepted in the two-timing method.

4. The justification of the solutions of derived equations can be performed in the same way as in other cases of use of two-timing method.

5. It is surprising, that any treatment of averaged flows driven by oscillating body forces is unknown in the literature.

6. We believe that the presented form of equations can be generalised, specified and used in such high-impact areas as turbo-machinery, medicine, biological fluid dynamics, \emph{etc.}, where the use of various `effective body forces' represents one of actively used modelling approaches.

This research is partially supported by
the grant IG/SCI/DOMS/16/13 from Sultan Qaboos University, Oman.

\end{document}